\documentclass[%
 reprint,
amsmath, amssymb,
aps,
]{revtex4-2}

\usepackage{graphicx}% Include figure files
\usepackage{dcolumn}% Align table columns on decimal point
\usepackage{bm}% bold math
\usepackage{lineno}

\usepackage{amsmath}
\usepackage{amsthm}
\usepackage{float}
\usepackage{upgreek}
\usepackage{booktabs}
\usepackage{siunitx}
\usepackage[OT1]{fontenc}
\begin{document}

\preprint{APS/123-QED}

\title{Deep insight into charge equilibration and the effects on producing neutron-rich isotopes around $N = 126$ in the multinucleon transfer reactions}% Force line breaks with \\
%\thanks{A footnote to the article title}%

\author{Zehong Liao$^{1}$}
\author{Long Zhu$^{1}$
\footnote{Corresponding author: zhulong@mail.sysu.edu.cn}}
\author{Jun Su$^{1}$}
\author{Cheng Li$^{2}$}

\affiliation{%
$^{1}$Sino-French Institute of Nuclear Engineering and Technology, Sun Yat-sen University, Zhuhai 519082, China\\
$^{2}$School of Physics and Electrical Engineering, Anyang Normal University, Anyang 455000, Henan, China\\
}%

\date{\today}% It is always \today, today,

\begin{abstract}
The dynamics of the charge equilibration (CE) and the effects on the production of the neutron-rich isotopes around $N = 126$ in multinucleon transfer reactions are still not well understood. In this work, we investigate the mechanism of the CE from different viewpoints by using the extended version of the dinuclear system model (DNS-sysu) and the improved quantum molecular dynamics (ImQMD) model.
From the macroscopic and microscopic dynamical viewpoints, we find incomplete CE for the mass asymmetry reaction systems even
in the very deep collisions, and the behavior of ``inverse CE" that the tendency of the fragments is away from the $N/Z$ value of the compound system in the reaction $^{140}$Xe + $^{198}$Pt. Unlike the slow process presented in the ImQMD model, the behavior of fast equilibration with the characteristic time $\sim$ 0.52 zs is obtained based on the DNS-sysu model, which is consistent with the experimental data. By performing the systematic calculation, the correlation between the CE and the mass asymmetry of the reaction systems is clarified, which not only accounts for the observed intriguing phenomena of the CE but also sheds light on the optimal combinations for producing the neutron-rich isotopes around $N = 126$.
\end{abstract}

%\keywords{Suggested keywords}%Use showkeys class option if keyword
                              %display desired
\maketitle

%\tableofcontents

\section{\label{Introduction}Introduction}
Because of the promised potential for producing exotic nuclei \cite{zhu2020theoretical,zhang2018production,corradi2013multinucleon,loveland2019synthesis,adamian2020extend,sekizawa2019tdhf,li2016multinucleon,wu2019microscopic,krolas2003gamma,mignerey1980dependence,wen2020multinucleon,zagrebaev2011production,2020saikomnt}, the multinucleon transfer (MNT) process in the massive nuclei collisions has attracted a lot of attention~\cite{volkov1978DIC}. In particular, for the producing neutron-rich heavy nuclei around $N=126$, which are not only interesting in the nuclear structure but also attribute significantly in the understanding of the so called r-process ~\cite{grawe2007nuclear}, the great advantages of the cross sections in the MNT reactions have been demonstrated \cite{WOS:000259377500026,watanabe2015pathway,barrett2015xe,kozulin2012mass,heinz2014deep,beliuskina2014synthesis}.

The deep inelastic collisions (DIC) between nuclei in contact result in the profound reconstruction of the initial nuclei with the incident energy dissipation. The DIC at low energies shows great potential to study the charge equilibration (CE) by tracking the differential motion of protons and neutrons~\cite{mcintosh2019interplay}. The equilibration of neutron-to-proton ratios ($N/Z$) that takes place at the early stage of the collision has been noticed~\cite{kratz1977charge,hernandez1981quantal,PhysRevLett.118.062501,2019transport,SIMENEL2012607,Ayik2021smf,sekizawa2020,saiko2020role}. Based on the time-dependent Hartree-Fock (TDHF), Simenel \emph{et al}. compared the timescale for the CE and other processes, and indicated that the CE should not be the sole dissipation way   \cite{simenel2020timescales}. On the other hand, several works have suggested that the CE is related to the quantum features, e.g., the giant dipole resonance mode (GDR) ~\cite{simenel2001quantum,baran2001collective,prl2010Suppression}. However, the interplay between equilibration and dissipation in quantum systems is still not well understood.

The difference in the $N/Z$ of reaction partners influences the nucleon transfer direction and then the formation probabilities of exotic fragments in the MNT process \cite{WOS:000417491600001,bao2018influence,zhu2017theoretical}. Moreover, inspired by the broad range of $N/Z$ values, the developments of the radioactive beam facilities around the world, such as FRIB (USA) \cite{frib2017facility}, RIKEN (Japan) \cite{2010riken}, SPIRAL2 (France) \cite{2011spiral2}, and HIAF (China) \cite{yang2013hiaf}, provide a great opportunity for getting deep insight into the mechanism of the CE and the favorable projectiles for producing exotic isotopes. The better understanding of CE will lead to more reliable predictions for producing neutron-rich isotopes around $N=126$ in the MNT reactions.

In this work, the mechanism of the CE in the MNT reactions induced by Xe isotopes is studied by using the extended version of the dinuclear system model (DNS-sysu) and the improved quantum molecular dynamics (ImQMD) model. To interpret the CE behaviors and provide the essential information for producing $N=126$ neutron-rich isotopes in the MNT process, the correlations between the CE mode and the mass asymmetry $\eta$ (= (A$_{\textrm{target}}$ - A$_{\textrm{projectile}}$)/(A$_{\textrm{target}}$ + A$_{\textrm{projectile}}$)) of the combinations are investigated in the systematic study.

\section{\label{Model}Models}
The DNS-sysu model has been successfully used to describe the MNT reactions~\cite{zhu2019Uadvantages,zhu2021shell,zhu2021unified}. In the DNS-sysu model, (1) the master equation is extended by introducing the dynamical deformation degree of freedom; (2) the temperature dependence of potential energy surface (PES) is involved; (3) the unified description of fusion and MNT processes is achieved with the extension of the fusion concept~\cite{zhu2021unified}. In the DNS-sysu model, the production cross sections of the primary products with proton number $Z_{1}$ and neutron number $N_{1}$ can be calculated as follows:

\begin{flalign}
  \begin{split}
    \sigma_{\textrm{pr}}(Z_{1},N_{1},E_{\textrm{c.m.}})=\frac{\pi\hbar^{2}}{2\mu E_{\textrm{c.m.}}}\sum_{J=0}^{J_{\textrm{max}}}(2J+1)T_{\textrm{cap}}(J,E_{\textrm{c.m.}})\times\\
    \sum_{\beta_{2}} P(Z_{1},N_{1},\beta_{2},J,E_{\textrm{c.m.}},\tau_{\textrm{int}})\times W_{\textrm{sur}}(Z_{1},N_{1},J,E^{*}),
  \end{split}
\end{flalign}
where $T_{\mathrm{cap}}$ means the capture probability which is calculated with the Hill-Wheeler formula in the consideration of Coulomb barrier distribution. For the heavy systems without potential pockets, the value of $T_{\mathrm{cap}}$ is estimated as 1 if the incident energy is above the interaction potential at the contact configuration. $P$ is the distribution probability of the primary fragments with proton number $Z_{1}$ and neutron number $N_{1}$. $\beta_{2}$ is the dynamical deformation parameter of the DNS. $W_{\textrm{sur}}$ is the de-excitation probability of the excited primary fragments. The contact time $\tau_{\textrm{int}}$ can be calculated by the deflection function method \cite{li1983distribution,wolschin1978analysis}. 

The distribution probability $P$ can be obtained by solving the master equations numerically with the corresponding potential energy surface, which can be written as 
\cite{zhu2018prc}:
\begin{flalign}
\begin{split}\label{master}
&\frac{dP(Z_{1}, N_{1}, \beta_{2}, J, t)}{dt}\\
&=\sum_{Z_{1}^{'}}W_{Z_{1}, N_{1}, \beta_{2}; Z_{1}^{'}, N_{1}, \beta_{2}}(t)[d_{Z_{1}, N_{1}, \beta_{2}}P(Z_{1}^{'}, N_{1}, \beta_{2}, J, t)\\
&-d_{Z_{1}^{'}, N_{1}, \beta_{2}}P(Z_{1}, N_{1}, \beta_{2}, J, t)]\\
&+\sum_{N_{1}^{'}}W_{Z_{1}, N_{1}, \beta_{2};Z_{1}, N_{1}^{'}, \beta_{2}}(t)[d_{Z_{1}, N_{1}, \beta_{2}}P(Z_{1}, N_{1}^{'}, \beta_{2}, J, t)\\
&-d_{Z_{1}, N_{1}^{'}, \beta_{2}}P(Z_{1}, N_{1}, \beta_{2}, J, t)]\\
&+\sum_{\beta_{2}^{'}}W_{Z_{1}, N_{1}, \beta_{2};Z_{1}, N_{1}, \beta_{2}^{'}}(t)[d_{Z_{1}, N_{1}, \beta_{2}}P(Z_{1}, N_{1}, \beta_{2}^{'}, J, t)\\
&-d_{Z_{1}, N_{1}, \beta_{2}^{'}}P(Z_{1}, N_{1}, \beta_{2}, J, t)],
\end{split}
\end{flalign}
where $W_{Z_{1}, N_{1}, \beta_{2};Z_{1}^{'}, N_{1}, \beta_{2}}$ denotes the mean transition probability from the channel ($Z_{1}$, $N_{1}$, $\beta_{2}$) to ($Z_{1}^{'}$, $N_{1}$, $\beta_{2}$), which is similar to $N_{1}$ and $\beta_{2}$. $d_{Z_{1}, N_{1}, \beta_{2}}$ is the microscopic dimension (the number of channels) corresponding to the macroscopic state ($Z_{1}$, $N_{1}$, $\beta_{2}$). For the degrees of freedom of charge and neutron number, the sum is taken over all possible proton and neutron numbers that fragment 1 may take, but only one nucleon transfer is considered in the model ($Z_{1}^{'} = Z_{1} \pm 1$; $N_{1}^{'} = N_{1} \pm 1$). For the $\beta_{2}$, we take the range of -0.5 to 0.5. The evolution step length is 0.01. The transition probability is related to the local excitation energy \cite{ayik1976microscopic}, in which the memory time is $0.25\tau_{\textrm{0}}/\mathit{A}$. Here, $\tau _{0} \equiv 2\pi \hbar/(1 \textrm{MeV})\approx 4\cdot 10^{-21}$\textrm{sec} , $\mathit{A}$ means the total nucleon number of the reaction \cite{norenberg1975quantum}.

The potential energy surface (PES) can be written as:
\begin{flalign}\label{PES}
\begin{aligned}
  &U\left(Z_{1}, N_{1}, \beta_{2}, J, r=R_{\text {cont }}\right)= \Delta\left(Z_{1}, N_{1}\right)+\Delta\left(Z_{2}, N_{2}\right) \\
  &+V \left(Z_{1}, N_{1}, \beta_{2}, J, r=R_{\text {cont}}\right)+\frac{1}{2} C_{1}\left(\delta \beta_{2}^{1}\right)^{2}+\frac{1}{2} C_{2}\left(\delta \beta_{2}^{2}\right)^{2}.
  \end{aligned}
\end{flalign}
Here, $\Delta\left(Z_{i}, N_{i}\right)$ ($i$ =1,2) is the mass excess of the $i$th fragment \cite{zhu2018shell}. The last two terms are the deformation energies which can be calculated using the methods shown in Refs. \cite{zhu2018prc,zagrebaev2007shell}. $R_{\text {cont}}$ is the position where the nucleon transfer process takes place \cite{zhu2019Uadvantages}.

The effective nucleus-nucleus interaction potential $V$ consists of the long-range Coulomb repulsive potential, the attractive short-range nuclear potential, and the centrifugal potential:
\begin{flalign}
\begin{aligned}
  V\left(Z_{1}, N_{1}, \beta_{2}, J, r \right) &=V_{\mathrm{N}}\left(Z_{1}, N_{1}, \beta_{2}, r\right) \\
  &+V_{\mathrm{C}}\left(Z_{1}, N_{1}, \beta_{2}, r\right)+\frac{(J \hbar)^{2}}{2 \zeta_{\text {rel }}}.
\end{aligned}
\end{flalign}

where, $\zeta_{\text {rel}}$ is the moment of inertia for the relative motion of the DNS. More detailed description of Coulomb potential $V_{C}$ and nuclear potential $V_{N}$ can be seen in Refs. \cite{adamian1996effective,wong1973interaction}.
To systematically study the evolution of the CE, all reactions in this work are calculated and studied at 1.2 times interaction potential energies at the contact positions.
%ImQMD

The ImQMD model \cite{2002wangqmd} is an improved version of the quantum molecular dynamics model \cite{qmd1991aichelin}. In this version, the Hamiltonian of the system is written as the sum of the kinetic energy $T = \sum_{i}\frac{p_{i}^{2} }{2m_i} $ and effective interaction potential energy.
\begin{flalign}
\begin{split}\label{hamiltonian}
  H = T + U_{\text {Coul}} + U_{\text {loc}} .
\end{split}
\end{flalign}
Here, $U_{\text {Coul}}$ is the Coulomb interaction potential energy
\begin{flalign}
\begin{split}
U_{\text {Coul}}=& \frac{1}{2} \iint \rho_{p}(\mathbf{r}) \frac{e^{2}}{\left|\mathbf{r}-\mathbf{r}^{\prime}\right|} \rho_{p}\left(\mathbf{r}^{\prime}\right) d \mathbf{r} d \mathbf{r}^{\prime} \\
&-e^{2} \frac{3}{4}\left(\frac{3}{\pi}\right)^{1 / 3} \int \rho_{p}^{4 / 3} d \mathbf{r}.
\end{split}
\end{flalign}
with $\rho_{p}$ the density distribution of protons of the system. $U_{\text{loc}}$ is the nuclear interaction potential energy, which is obtained from the integration of the Skyrme energy density functional $U_{\text{loc}} = \int V_{\text{loc}} (\mathbf{r})\mathrm{d}\mathbf{r}$ without the spin-orbit term. The nuclear interaction potential density $V_{\text{loc}}$ can be written as
\begin{flalign}
\begin{split}
V_{\mathrm{loc}}=& \frac{\alpha}{2} \frac{\rho^{2}}{\rho_{0}}+\frac{\beta}{\gamma+1} \frac{\rho^{\gamma+1}}{\rho_{0}^{\gamma}}+\frac{g_{\mathrm{sur}}}{2 \rho_{0}}(\nabla \rho)^{2}+\\
& \frac{C_{\mathrm{s}}}{2 \rho_{0}}\left[\rho^{2}-\kappa_{\mathrm{s}}(\nabla \rho)^{2}\right] \delta^{2}+g_{\tau} \frac{\rho^{\eta+1}}{\rho_{0}^{\eta}},
\end{split}
\end{flalign}
where, $\delta$ is the isospin asymmetry. The parameters are shown in Table~\ref{tab:table1}.
The density distribution in the coordinate space $\rho(\mathbf{r})$ is given by
\begin{flalign}
\begin{split}
  \rho(\mathbf{r})=\sum_{i} \frac{1}{\left(2 \pi \sigma_{r}^{2}\right)^{3 / 2}} \exp \left[-\frac{\left(\mathbf{r}-\mathbf{r}_{i}\right)^{2}}{2 \sigma_{r}^{2}}\right].
\end{split}
\end{flalign}

Where, $\sigma_{r}$ is the Gaussian wave pocket variation. The IQ2 parameter sets (see in Table~\ref{tab:table1}) are adopted in this work. The fermionic nature in ImQMD is treated as the method proposed by Papa \emph{et al} \cite{papaprc2001}. In the MNT reactions, the total kinetic energy-mass distributions of the primary binary fragments in different contact time ranges \cite{liplb2020} or impact parameters \cite{liplb2018} can serve as an ethical basis for classifying different channels. The exotic fragments or specific objective nuclei could be produced in the channels of deep-inelastic collisions, quasi-fission, and quasi-elastic collisions. Therefore, the events from central collisions to grazing are considered in this work.
\begin{table}[h]
\footnotesize
\caption{\label{tab:table1} The model parameters (IQ2) adopted in this work}
\begin{center}
\setlength{\tabcolsep}{1.2mm}
\begin{tabular}{ccccccccc}
\hline
$\alpha$ & $\beta$ & $\gamma$  & $g_{sur}$ & $g_{\tau}$ & $\eta$ & $C_{s}$ & $\kappa_{s}$ & $\rho_{0}$\\
$\text{MeV}$  &  $\text{MeV}$   &  & ($\text{MeV$\cdot$ fm$^{2}$}$)  & $\text{MeV}$ &  & $\text{MeV}$ & $\text{fm$^{2}$}$ & $\text{fm$^{-3}$}$\\
\hline
-356  &  303  &  7/6 &  7.0  &  12.5 & 2/3 & 32.0 & 0.08 & 0.165 \\
\hline
\end{tabular}
\end{center}
\end{table}
To get the production cross sections of the final products after the de-excitation process, the code GEMINI++ \cite{charity1988systematics} is used to treat the de-excitation process of primary fragments. Subsequent de-excitation cascades of the excited fragments via emission of light particles (neutron, proton, and $\alpha$, etc) and gamma-rays competing with the fission process leads to the final mass distribution of the reaction products.

\section{\label{Results and discussions}Results and discussions}
%picture one and text
\begin{figure}[t]
\begin{center}
\includegraphics[width=9cm,height=9cm,angle=0]{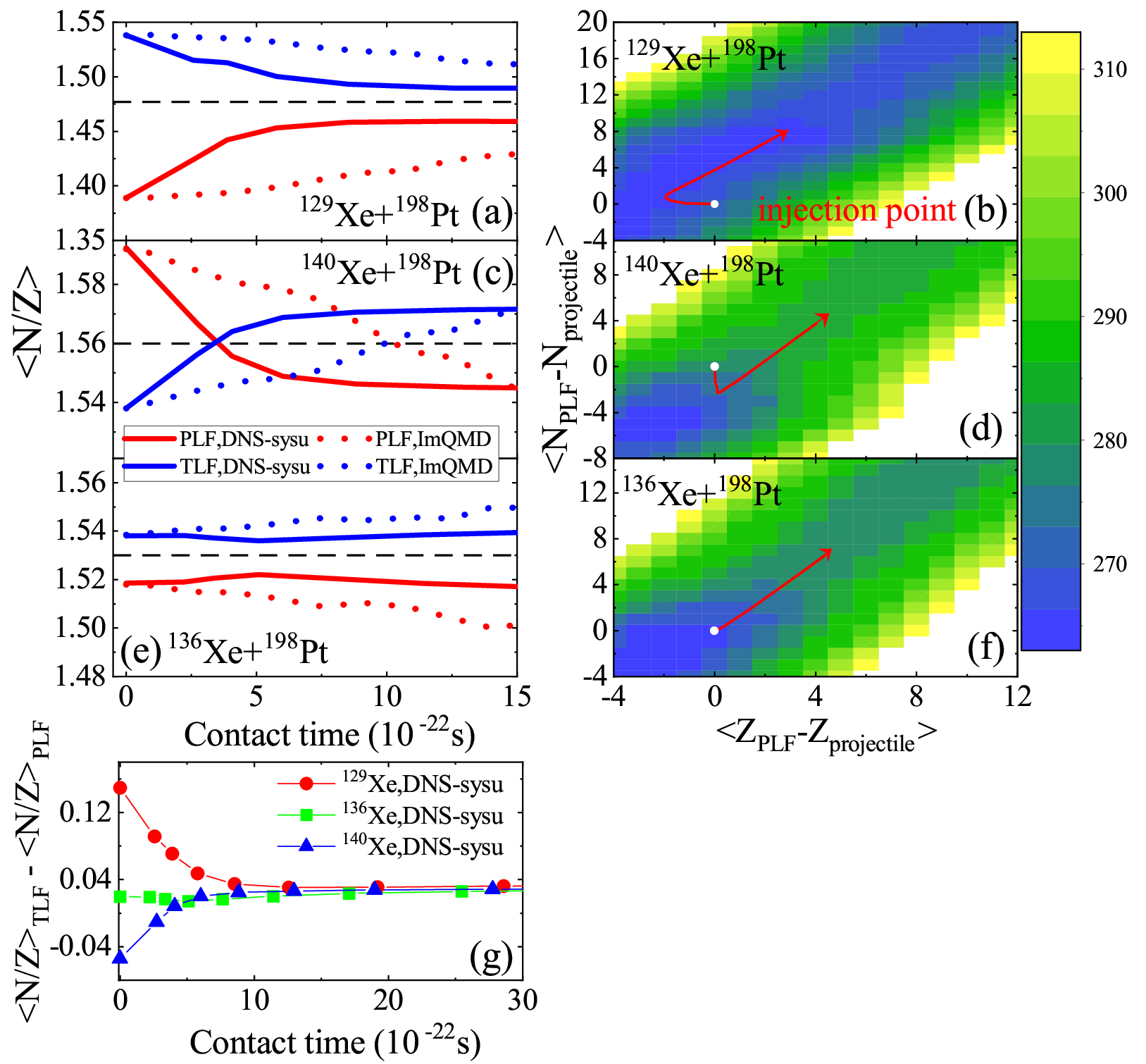}
\caption{
(Color online). Left panels: Calculated average $N/Z$ values of the PLF and TLF in the
reactions $^{129}$Xe + $^{198}$Pt (a), $^{140}$Xe +$^{198}$Pt (c), and $^{136}$Xe + $^{198}$Pt (e) as a function of the contact time within the DNS-sysu and ImQMD models. The $N/Z$ values of compound systems are also shown with the black horizontal dashed lines. Right panels: Contour plots of the PES (in MeV) with the drift trajectories of the first moments of PLF distributions in a $\langle Z_{\textrm{PLF}} - Z_{\textrm{Projectile}} \rangle$, $\langle N_{\textrm{PLF}} - N_{\textrm{Projectile}} \rangle$ plane for the reactions $^{129}$Xe + $^{198}$Pt (b), $^{140}$Xe +$^{198}$Pt (d), and $^{136}$Xe + $^{198}$Pt (f). (g): The discrepancies of average $N/Z$ values between the TLF and PLF as a function of the interaction time calculated in the DNS-sysu model for the reactions $^{129}$Xe, $^{136}$Xe, $^{140}$Xe + $^{198}$Pt. The incident energies $E_{\textrm{c.m.}}= $ 470, 476, and 466 MeV for the reactions induced by $^{129,136,140}$Xe. }\label{picture1}\end{center}\end{figure}
To clarify the $N/Z$ asymmetry effect in MNT reactions, Fig.~\ref{picture1}(a) shows the average $N/Z$ values of the projectile-like fragments (PLF) and target-like fragments (TLF) as a function of the contact time in the reaction $^{129}$Xe + $^{198}$Pt within the framework of the DNS-sysu model. The large discrepancy of $N/Z$ values between $^{129}$Xe and $^{198}$Pt, e.g., $N/Z$ = 1.39 for $^{129}$Xe, $N/Z$ = 1.54 for $^{198}$Pt, is expected to confirm the presence of CE behaviors. One interesting behavior can be seen that the average $N/Z$ values of the PLF and TLF trend toward the $N/Z$ value of the compound system ($N/Z$ = 1.48) but do not lead to being identical even in very deep collisions. We also show the results from the ImQMD model. The incomplete CE is also noticed. Besides, unlike the results from the DNS-sysu model, the CE takes place at the whole collision stage gradually.

In principle, the MNT process can be described as the reaction upon the so-called PES, where the dynamical evolution of a dinuclear system can be treated as the process of exchanging the independent-particles between the nuclei~\cite{freiesleben1984nz}. Plotted in Fig.~\ref{picture1}(b), (d), and (f), the PES can be defined as the contour of  the first moments of the PLF distributions, i.e., $\langle Z_{\textrm{PLF}} - Z_{\textrm{Projectile}} \rangle$, $\langle N_{\textrm{PLF}} - N_{\textrm{Projectile}} \rangle$. Taking the reaction $^{129}$Xe + $^{198}$Pt in Fig.~\ref{picture1}(b) as an example, the trajectory starts from the injection point and relaxes to the valley of the PES driven by the PES gradient (See the red line in Fig.~\ref{picture1}(b)), which indicates that the projectile tends to lose protons and absorb neutrons from the target.

In Fig. \ref{picture1}(c), we also show the results of the reaction $^{140}$Xe + $^{198}$Pt. The $N/Z$ value of the projectile $^{140}$Xe is 1.59, which is larger than that of the target. However, unlike the reaction $^{129}$Xe + $^{198}$Pt, the inverse relationship ($N/Z$$_{(\textrm{PLF})}$ $<$ $N/Z$$_{(\textrm{TLF})}$) is noticed during the evolution. The similar behavior is also presented from the ImQMD calculations, except the crossover appears at a quite delayed contact time. The transfer of neutrons from $^{140}$Xe to $^{198}$Pt is promoted, due to negative values of $\Delta U$ in the neutron stripping channels, such as $\Delta U_{\textrm{-1n}}$ = -2.6 MeV, $\Delta U_{\textrm{-2n}}$ = -4.8 MeV, and $\Delta U_{\textrm{-3n}}$ = -6.8 MeV. Here, $\Delta U_{- \mathit{x} n} [=U(Z_{\textrm{P}}, N_{\textrm{P}}- \mathit{x}, \beta_{2}=0, J=0, r=R_{\text {cont }})-U(Z_{\textrm{P}}, N_{\textrm{P}}, \beta_{2}=0, J=0, r=R_{\text {cont }})]$ represents the driving potential needed to be overcome in the neutron transfer process. The detailed description of $U$ is shown in Eq. (\ref {PES}).  The system $^{140}$Xe + $^{198}$Pt with negative values of $\Delta U_{-\mathit{x}n}$ tends to enhance the average $N/Z$ values of the TLF and decrease those of the PLF. 

Compared with the initial entrance channel, the behavior called ``inverse CE", producing the fragments with the average $N/Z$ values farther from the $N/Z$ value of the compound system ($N/Z$ = 1.56), is noticed. This is a strong indication that the underlying mechanism behind the CE does not only depend on the $N/Z$ asymmetry in the entrance channel.

Furthermore, considering the different initial $N/Z$ values for the projectile and target in the reaction $^{136}$Xe + $^{198}$Pt ($N/Z$ = 1.52 for $^{136}$Xe, $N/Z$ = 1.54 for $^{198}$Pt), the behavior of CE should also be noticed on some levels. However, as shown in Fig. \ref{picture1}(e), the almost flat variation of the average $N/Z$ values with the interaction time is presented in both the DNS-sysu and ImQMD model calculations. The PES of the reaction $^{136}$Xe + $^{198}$Pt is shown in Fig. \ref{picture1}(f).Since the trajectory sticks to the bottom of the valley, the weak variation of the average $N/Z$ values can be explained.

To compare the above systems, in Fig. \ref{picture1}(g), we show the differences between the average $N/Z$ values of the TLF and PLF ($\langle N/Z\rangle_{\textrm{TLF}}$-$\langle N/Z\rangle_{\textrm{PLF}}$) as a function of the interaction time calculated in the DNS-sysu model for the reactions $^{129,136,140}$Xe + $^{198}$Pt. It is interesting to see that for all three systems the values of $\langle N/Z\rangle_{\textrm{TLF}}$-$\langle N/Z\rangle_{\textrm{PLF}}$ evolve to a saturation value of 0.04, rather than 0.

%picture two and text
\begin{figure}[t]
\begin{center}
\includegraphics[width=8cm,angle=0]{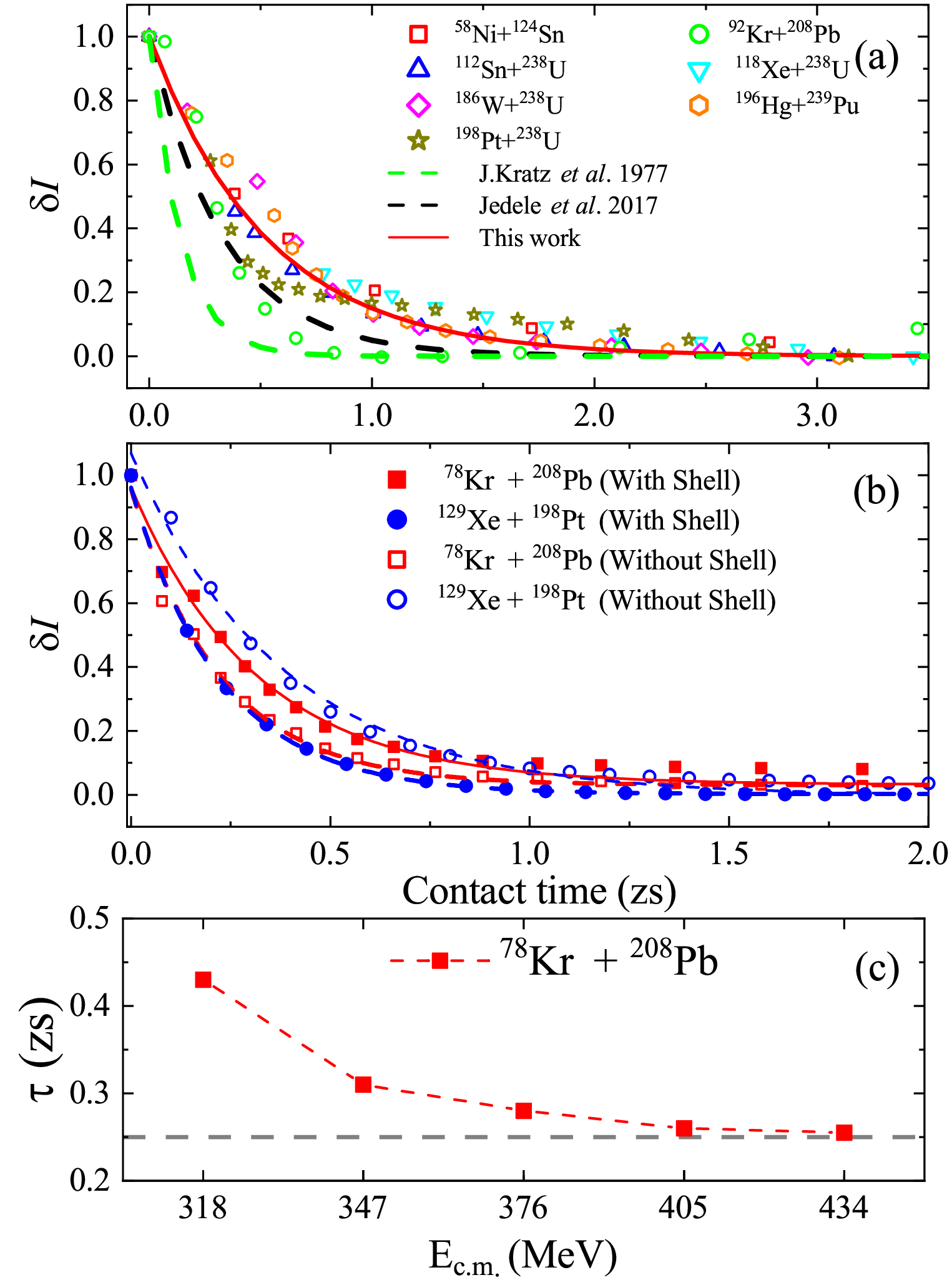}
\caption{(Color online). (a) The evolution degree of the CE ($\delta$$I(t)$) as a function of contact time from DNS-sysu calculation of the reactions $^{58}$Ni + $^{124}$Sn, $^{92}$Kr + $^{208}$Pb, $^{112}$Sn + $^{238}$U, $^{118}$Xe + $^{238}$U, $^{186}$W + $^{238}$U, $^{196}$Hg + $^{239}$Pu, and $^{198}$Pt + $^{238}$U. The green and black dashed lines show the expected equilibration assuming the rate constant of 0.14 zs reported experimentally by Kratz \emph{et al}. \cite{kratz1977charge} and 0.3 zs reported experimentally by Jedele \emph{et al}. \cite{PhysRevLett.118.062501}, respectively.
(b) Evolution of $\delta I$ as a function of contact time from DNS-sysu calculations of $^{78}$Kr + $^{208}$Pb (squares) and $^{129}$Xe + $^{198}$Pt (circles). The solid symbols and open symbols denote the results with and without shell corrections, respectively. The solid line and dashed line denote the expected fitting results with and without shell corrections, respectively. (c) The CE characteristic time $\tau$ as a function of the incident energy for the reaction $^{78}$Kr + $^{208}$Pb calculated in the DNS-sysu model. The horizontal dashed line is used to guide the eye.}
\label{picture2}\end{center}\end{figure}
From the equilibration perspective timescale, the CE process is believed to be the fastest one among all the equilibration modes~\cite{rehm1979time}. To compare the equilibration timescale of various systems, we introduce a general ``normalized" observable $\delta I(t)$ = ($I(t)$ - $I_{\infty}$) / ($I_{0}$ - $I_{\infty}$) \cite{simenel2020timescales}. Here, $I_{0}$ and $I_{\infty}$ respectively denote the initial and expected saturation $N/Z$ values of the projectile. During the evolution, it is reasonable to see that the $\delta I$ decays from 1 to 0. The so-called characteristic time $\tau$ can be adopted to characterize the CE process, which is the parameter denoting the decay of $\delta I$ following $\delta I=y_{0}+A_{0} \exp (-t/\tau)$.

In Fig. \ref{picture2}(a), several systems are calculated to present a comprehensive timescale of the CE in this work. The incident energy of $E_{\textrm{c.m.}}$ = $1.2V(r=R_{\textrm{cont}})$ is used for each reaction. The values of $V(r=R_{\textrm{cont}})$ are 143, 276, 423, 439, 573, 650, and $\SI{618}{MeV}$, respectively, for the reactions $^{58}$Ni + $^{124}$Sn, $^{92}$Kr + $^{208}$Pb, $^{112}$Sn + $^{238}$U, $^{118}$Xe + $^{238}$U, $^{186}$W + $^{238}$U, $^{196}$Hg + $^{239}$Pu, and $^{198}$Pt + $^{238}$U. Despite the fluctuations, all systems exhibit a similar and fast decay pattern. By fitting the data in the above reactions, 0.52 zs of the characteristic time $\tau$ is obtained in this work (red line), which is in the same order of magnitude as the experimental data ($\tau \sim \SI{0.3}{zs}$) in Ref. \cite{PhysRevLett.118.062501}, the measurement ($\tau \sim \SI{0.14}{zs}$) by Kratz \emph{et al}. ~\cite{kratz1977charge}, and $\SI{0.5}{zs}$ from the TDHF calculations~\cite{2019transport}. Applying the microscopic stochastic mean-field approach ~\cite{Ayik2021smf}, Ayik \emph{et al}. performed a similar CE calculation. The system reaches equilibrium after $\SI{1}{zs}$ from first touching, which is also consistent with the result in this work. 
  
Furthermore, seen from Fig.~\ref{picture1}, the average $N/Z$ values of PLF or TLF, estimated based on the DNS-sysu and ImQMD models, show quite a different trend with contact time elapsed. The slow CE process shown in the ImQMD model might be due to the absence of the spin-orbit coupling in the interaction and the details of the Pauli blocking treatment \cite{DONANGELO199758, PhysRevC.90.064612}.

The shell effects on the CE are also investigated in the reactions $^{78}$Kr + $^{208}$Pb and $^{129}$Xe + $^{198}$Pt within the DNS-sysu model at $E_{\textrm{c.m.}}=347$ and $\SI{470}{MeV}$, respectively. Note that the characteristic time extracted from the reaction $^{78}$Kr + $^{208}$Pb with shell corrections ($\SI{0.31}{zs}$) is greater than that without shell corrections ($\SI{0.22}{zs}$). The delayed CE is because the nucleon transfer process would be inhibited due to the doubly magic $^{208}$Pt target. However, an enhancement of the shell effects in the equilibrium speed can be observed in the reaction $^{129}$Xe + $^{198}$Pt ($\tau_{\textrm{with-shell}} \sim \SI{0.27}{zs}$, $\tau_{\textrm{without-shell}} \sim \SI{0.38}{zs}$). This enhancement can be attributed to the attraction of the proton shell closure $Z = 82$ for transferring protons from the $^{129}$Xe to $^{198}$Pt. In addition, the phenomenon that $^{129}$Xe is preferably driven to lost protons by the PES gradient can also be seen clearly in Fig. \ref{picture1}(d). As we can see, for each reaction, our calculations demonstrate that the shell effects could influence the neutron and proton flow strongly. In Fig. \ref{picture2}(c), we show the CE characteristic time $\tau$ as a function of the incident energy. Interestingly, $\tau$ gradually decreases with the increase of incident energy and reaches saturation for the incident energy higher than 405 MeV (1.4$V(r=R_{\textrm{cont}})$). The high incident energy enhances the energy dissipating into the internal of the system, which promotes the process of equilibration. However, the time scale of equilibrium shows a limit, which can be roughly estimated by dividing the size of a nucleus by the speed at which isospin waves or nucleons propagate. We should point out, the absence of the microscopic characteristics prevents the DNS model from getting further investigating the CE process from the viewpoint of quantal fluctuation, which might play an important role in the fast CE process.

Due to the ``curvature" of the $\beta$-stability line, the heavy nuclei show the capability of possessing more neutrons than the light ones, which could affect the nucleon flow in the collisions. The CE would be strongly affected by the mass asymmetry of the reaction partners. Therefore, to clarify the interesting phenomena shown in Fig. \ref{picture1}, one conjecture can be made that the initial mass asymmetry of combinations also plays a significant role during the isospin transfer process.

\begin{figure}[t]
\begin{center}
\includegraphics[width=8cm,angle=0]{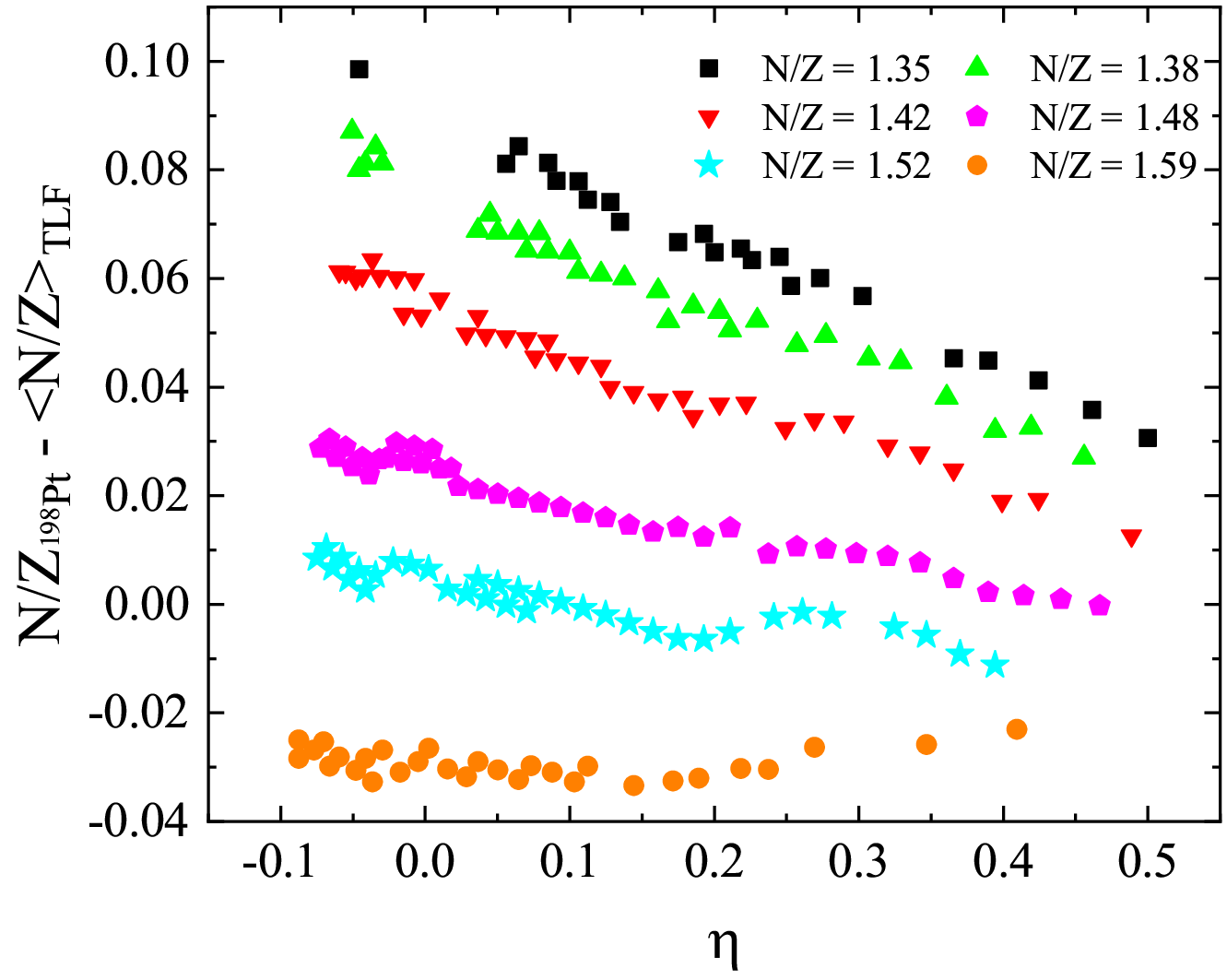}
\caption{(Color online). The values of $ N/Z_{^{198}\textrm{Pt}}$-$\langle N/Z\rangle$$_{\textrm{TLF}}$ in the MNT reactions with the target $^{198}$Pt as a function of the mass asymmetry of combinations and $N/Z$ values of the projectiles. The projectiles are selected based on the $N/Z$ values which are about 1.35, 1.38, 1.42, 1.48, 1.52, and 1.59.
}\label{picture3}\end{center}\end{figure}

To investigate the correlations between the CE and the mass asymmetry of the colliding combinations, the reactions with the projectiles spanning a broad range of masses and $N/Z$ values bombarding the target $^{198}$Pt are systematically studied. In Fig.~\ref{picture3}, we show the values of $N/Z_{^{198}\textrm{Pt}}-\langle N/Z_{\textrm{TLF}}\rangle$ as a function of the mass asymmetry of colliding combinations and the $N/Z$ values of the projectiles. Each symbol denotes a reaction system with specific $N/Z$ asymmetry and mass asymmetry. Despite the minor fluctuations, a correlation between the mass asymmetry and the $N/Z$ asymmetry is clearly shown in influencing the neutron richness of the TLF.
%picture three and text

As shown in Fig. \ref{picture3}, for the case of $N/Z=1.35$, the TLF neutron-richness strongly depends on the mass asymmetries of the colliding systems. One can see that the combinations with positive large values of $\eta$ show small discrepancies between the average $N/Z$ values of TLF and that of $^{198}$Pt, although the systems present large $N/Z$ asymmetries. The same behavior can be seen for the projectiles with $N/Z = 1.59$ (shown as brown circles) which is higher than 1.54 of $^{198}$Pt. This behavior testifies to the conjecture we made that the mass asymmetries of the reaction combinations strongly affect the CE process. Furthermore, it can be seen that the mass asymmetry dependence of CE is weakened for the $N/Z$ symmetric combinations. For example, for the reactions induced by projectiles with $N/Z$ values close to 1.52 (shown as blue stars), the weak variation of $N/Z_{^{198}\textrm{Pt}}-\langle N/Z_{\textrm{TLF}}\rangle$ values on the mass asymmetry can be seen. This is the main reason for the behavior shown in Fig. \ref{picture1}(e). Therefore, the intriguing phenomena shown in Fig. \ref{picture1}, such as incomplete CE and ``inverse CE" are mainly because of the correlation between the CE and the mass asymmetry. Here, we point out that the complete CE is hard to be reached in mass asymmetric systems.

%picture four and text
As stated above, the evolution of the CE can be influenced by the $N/Z$ asymmetry and the initial mass asymmetry. To further understand the effect of the CE process in MNT reactions for producing neutron-rich isotopes and provide guidance for selecting the favorable combinations, we define a relative isospin flow factor $K$, which can be written as:
\begin{flalign}
\begin{split}
K = \frac{(N/Z)_{\textrm{target}}  -  \langle N/Z\rangle_{\textrm{TLF}}  }{ (N/Z)_{\textrm{CN}} }.
\end{split}
\end{flalign}
The negative values of $K$ denote that there are advantages for producing the neutron-rich TLF. On the contrary, for the positive values of $K$, the corresponding projectiles could enhance the probabilities of producing the neutron-deficient TLF. For example, for the reactions $^{129,136,140}$Xe + $^{198}$Pt we studied above, the corresponding values of the factor $K$ are 0.034, 0, and 0.02, respectively. It is noticed that the value of the factor $K$ could sensitively present the isospin flow directions in different reactions.
\begin{figure}[t]
\begin{center}
\includegraphics[width=8cm,angle=0]{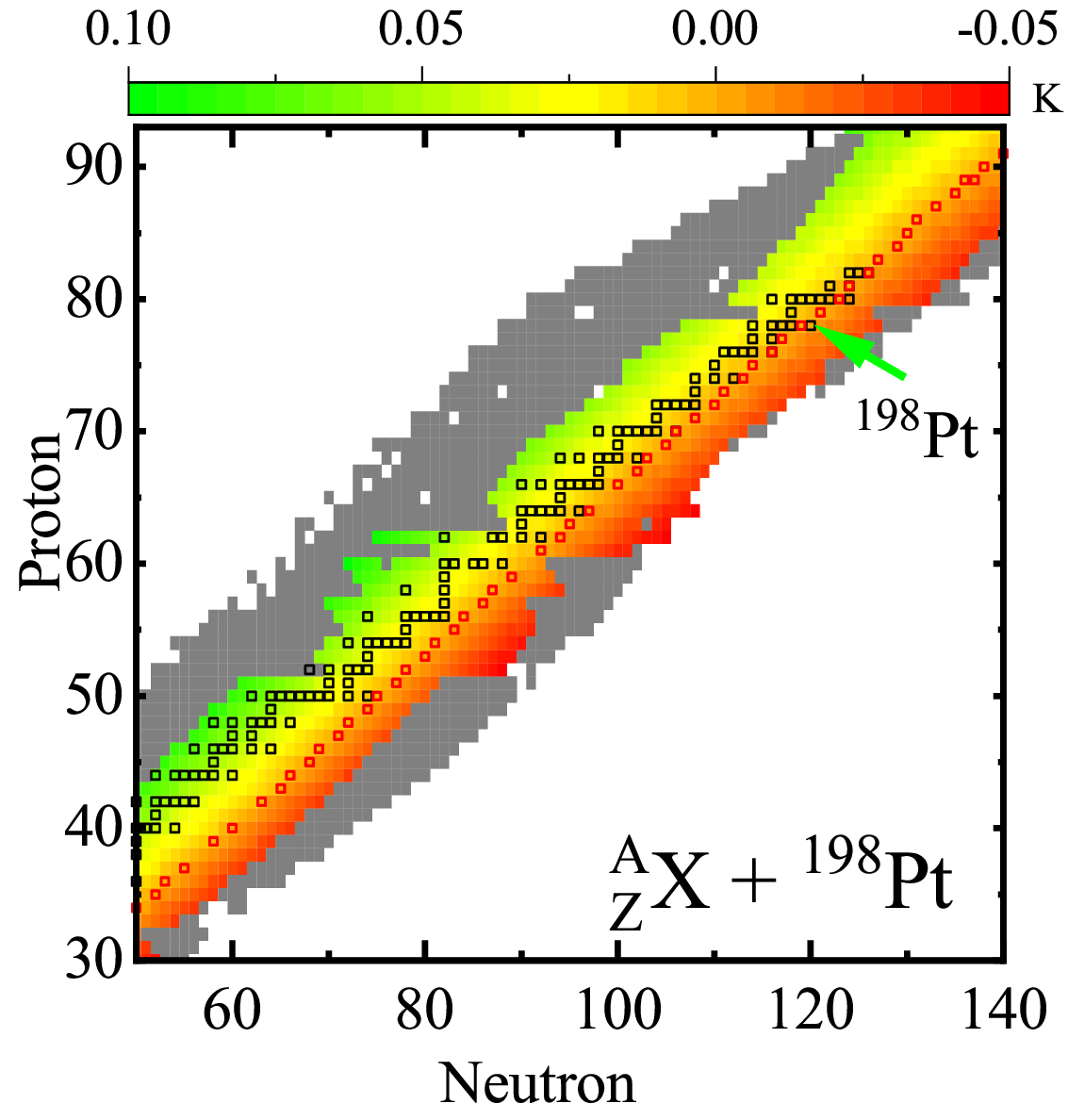}
\caption{ (Color online). The factor $K$ for the reactions based on $^{198}$Pt target as a function of the neutron and proton numbers of the projectiles. The black open squares show nuclide along the $\beta$-stability line. The red open squares denote the reactions with $K$ values that are approximately equal to 0.}\label{picture4}\end{center}\end{figure}
We systematically investigate the $K$ values in the reactions based on the $^{198}$Pt target. In Fig. \ref{picture4}, we show the distribution of the factor $K$ as a function of the neutron and proton numbers of the projectiles. The red open squares denote the reactions that $K$ values approximately equal to 0. In other words, for the reactions on the northwest of these red open squares in the chart, the average $N/Z$ values of the TLF are less than 1.54 of the $^{198}$Pt target. On the contrary, the average $N/Z$ values of TLF are greater than 1.54 of the $^{198}$Pt target for the reactions on the southeast of the red open squares. For the reactions farther away from these red open squares, the more neutron-rich or neutron-deficient TLF could be produced. The black open squares denote the reaction with the projectiles along the $\beta$-stability line. For the regions with $Z < 50$ and $N < 80$, we notice that the red open squares are located on the more neutron-rich side compared to the $\beta$-stability line. Hence, only extremely neutron-rich projectiles in this region could promote the production of the neutron-rich TLF. On the other hand, for the reactions with heavy projectiles, such as the mass symmetric reactions, it can be seen obviously that these red open squares are close to the $\beta$-stability line. It is worth mentioning that beams of nuclide around the $\beta$-stability line usually show higher intensities compared to those far from the $\beta$-stability. Therefore, reactions induced by heavy projectiles would be efficient for producing neutron-rich nuclei around $N = 126$.

%picture five and text
\begin{figure}[t]
\begin{center}
\includegraphics[width=8cm,angle=0]{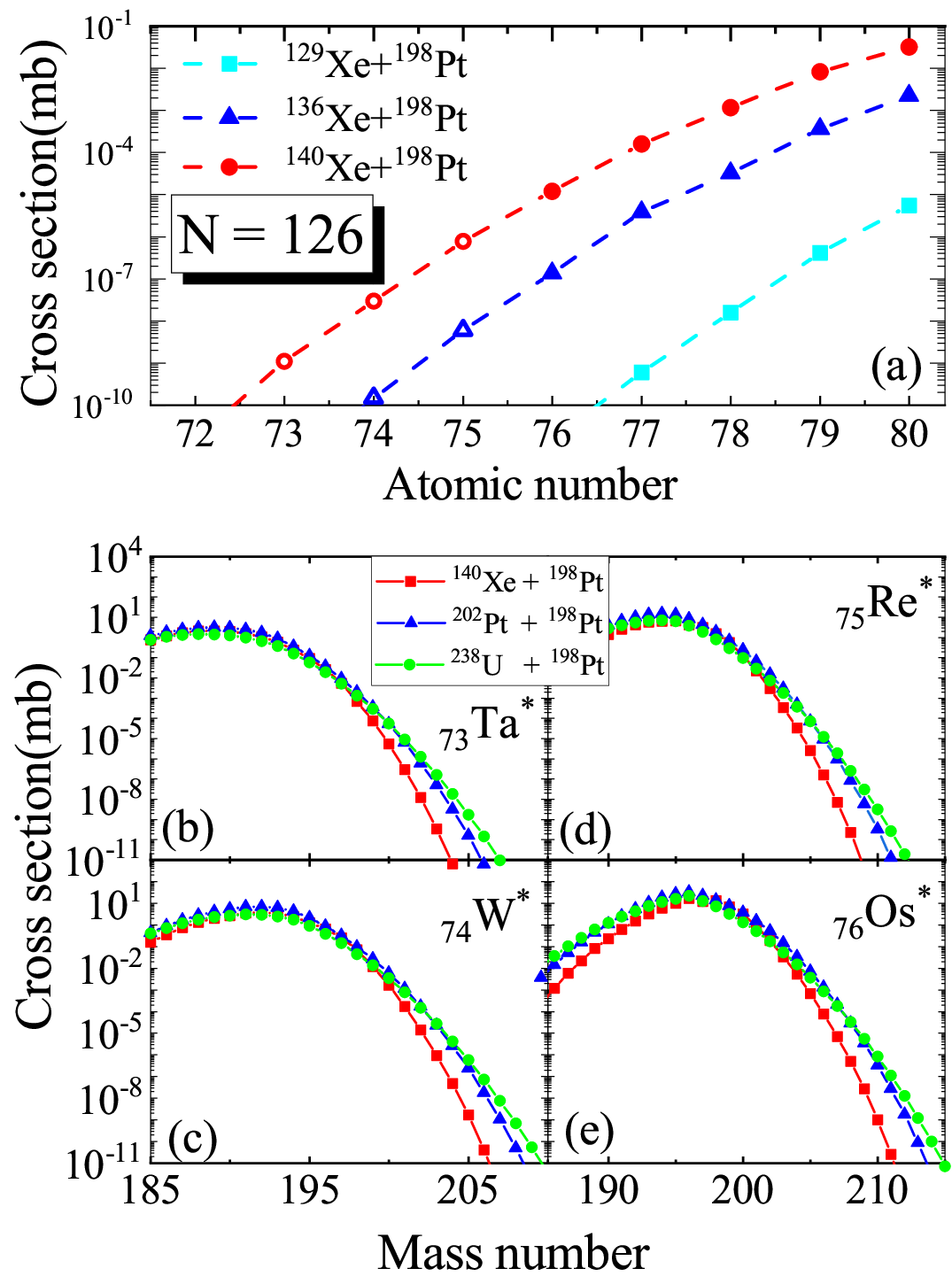}
\caption{(Color online). (a) Calculated production cross sections of the $N = 126$ isotones in the MNT reactions $^{129}$Xe, $^{136}$Xe, and $^{140}$Xe + $^{198}$Pt within the DNS-sysu model with GEMINI++ code. Open symbols denote unknown nuclei. (b)-(e) The calculated production cross sections of the primary fragments in the $^{140}$Xe + $^{198}$Pt, $^{202}$Pt + $^{198}$Pt, and $^{238}$U + $^{198}$Pt. }\label{picture5}\end{center}\end{figure}
In order to optimize the reaction combinations for producing unknown neutron-rich nuclei, the reactions $^{129}$Xe + $^{198}$Pt, $^{136}$Xe + $^{198}$Pt, $^{140}$Xe + $^{198}$Pt, $^{202}$Pt + $^{198}$Pt and $^{238}$U + $^{198}$Pt are investigated at incident energies of $E_{\textrm{c.m.}}= $ 470, 476, 466, 663 and 741 MeV, respectively. In Fig.~\ref{picture5}(a), we show the calculated production cross sections for $N = 126$ isotones with $Z < 80$ produced in the reactions $^{129,136,140}$Xe + $^{198}$Pt. As we expected, the reactions induced by projectiles with high neutron-richness ($N/Z$ $\sim$ 1.59 for $^{140}$Xe) show great advantages of cross sections for producing neutron-rich nuclei. This can be interpreted by the intense CE caused by the large $N/Z$ asymmetry. We also extract the cross sections of the primary fragments in the proton picking-up channels for the reactions $^{140}$Xe + $^{198}$Pt, $^{202}$Pt + $^{198}$Pt, and $^{238}$U + $^{198}$Pt shown in Fig.~\ref{picture5} (b)-(e). It can be seen that the production cross sections of the neutron-rich nuclei in the reaction induced by $^{140}$Xe are lower than those in the $^{202}$Pt and $^{238}$U induced ones, although $^{140}$Xe has an even larger $N/Z$ value. This is because the $^{140}$Xe + $^{198}$Pt reaction presents the positive large value of $\eta$ and the CE is inhibited, just like the correlation behaviors between $\eta$ and the CE process we discussed above. On the other side, the saturation values of $\langle N/Z\rangle_{\textrm{TLF}}-\langle N/Z\rangle_{\textrm{PLF}}$ for the reactions $^{238}$U + $^{198}$Pt and $^{202}$Pt + $^{198}$Pt are 0.01 and 0, respectively, which are close to the complete CE status. As we know, for the choice of optimal combinations, the beam intensity should be considered. Compared with the radioactive nuclide $^{202}$Pt and $^{140}$Xe, the $^{238}$U induced reaction is a better candidate for producing neutron-rich nuclei around $N = 126$.

\section{\label{Conclusions}Conclusions}
The mechanism of the CE is investigated within the DNS-sysu (macroscopic approach) and ImQMD (microscopic dynamical approach) models. It is found that the two models show different equilibration characteristic times. For the DNS-sysu model, it is noticed that the CE occurs at the early stage of the colliding. The equilibration characteristic time in the DNS-sysu model is about 0.52 zs in the same order of magnitude as the experimental data (0.3 zs) in Ref. \cite{PhysRevLett.118.062501}, and the TDHF calculations (0.5 zs). However, the ImQMD calculations show that the equilibration is a slow process and takes place in the whole colliding process. In addition, the shell effect on CE is studied based on the reactions $^{78}$Kr + $^{208}$Pb and $^{129}$Xe + $^{198}$Pt within the DNS-sysu model. And the obvious influence of shell closures and incident energy on CE characteristic time is noticed. In both the DNS-sysu and ImQMD calculations, it is found that (i) the complete CE is hard to be reached in the mass asymmetric reaction systems and (ii) the ``inverse CE" takes place in the reaction $^{140}$Xe + $^{198}$Pt. In this work,  by performing the systematic calculation, for the first time, the correlations between the CE and the mass asymmetry of the reaction systems are clarified, and the above intriguing CE behaviors are interpreted with superposition of the mass asymmetry and the charge asymmetry of the colliding combinations.

With the definition of the relative isospin flow factor $K$, the isospin flow in the $^{198}$Pt target-based reactions with different colliding partners are investigated systematically. It is found that the light projectiles are not good candidates for producing neutron-rich isotopes $N = 126$ in the consideration of the isospin flow and relatively low beam intensities. Besides, the results in this work suggest that the combination $^{238}$U + $^{198}$Pt is favorable for producing unknown $N = 126$ isotopes.

\section*{\label{ACKNOWLEDGMENTS}ACKNOWLEDGMENTS}L. Zhu thanks Dr. P.W. Wen, H.H. Wen, and S.L. Chen for their careful reading of the manuscript. This work was supported by the National Natural Science Foundation of China under Grants No. 12075327 and 11875328.

\bibliography{reference}
\end{document}